\documentstyle[prb,preprint,aps,epsf]{revtex}
\begin{document}
\draft
\title{Three Self-Consistent Kinematics in (1+1)D Special-Relativity}

\author{P. Fraundorf}

\address{
Department of Physics \& Astronomy
and
Center for Molecular Electronics, \\
University of Missouri-SL,
St. Louis MO 63121}

\date{\today}

\maketitle

\begin{abstract}
When introducing special relativity, an elegant
connection to familiar rules governing Galilean constant
acceleration can be made, by describing first the discovery at
high speeds that the clocks (as well as odometers) of different
travelers may proceed at different rates.  One may then show how
to parameterize any given interval of constant acceleration with
{\em either}:  Newtonian (low-velocity approximation) time, inertial
relativistic (unaccelerated observer) time, or traveler proper
(accelerated observer) time, by defining separate velocities for
each of these three kinematics as well.  Kinematic invariance
remains intact for proper acceleration since $m a_o = dE/dx$.  This
approach allows students to solve relativistic constant
acceleration problems {\em with the Newtonian equations}!  It also
points up the self-contained and special nature of the
accelerated-observer kinematic, with its frame-invariant time,
4-vector velocities which in traveler terms exceed Newtonian
values and the speed of light, and of course relativistic
momentum conservation.
\end{abstract}

; UMStL-CME-951101pf; gr-qc/9512012

\section{Introduction}

Paradigm changes in physics are often accompanied by a
transformation of language and a hardening of distinctions.
Thus the emphasis on acceleration as the second time
derivative of displacement {\em in teaching Newtonian
dynamics} has hardened \cite{HRW,Serway}, even as the
concept of relativistic mass and inertial frame acceleration is
being slowly replaced by Lorentz covariant definitions of
force and acceleration {\em in the teaching of special
relativity} \cite{Goldstein,French,Taylor}.  The intent of
this note is to assist the second process by showing that:
(i) introduction of three relativistic time-parameterizations,
{\em including the Newtonian one}, provides a natural bridge
for introductory physics students to some deep truths
about relativistic space-time BEFORE the comparatively
abstract concept of Lorentz transforms need be introduced,
and  (ii) there are two relativistic alternatives to the popular
inertial-frame perspective, which perspective sometimes makes
velocity-dependent mass a tempting simplification \cite{Serway,French},
questions the special-relativistic treatability of acceleration
\cite{Mainwaring}, and implies that travelers as well as spectators
are {\em bound} by inertial velocity light-speed limits \cite{Lagoute}.
We also show why students here have said that "without the 82
nano-roddenberry speed-limit, and with the right vehicle, it's amazing
how little of my time it would take for me to get from point A to point B!"

\section{Observations}

We consider in this paper only one-dimensional constant
acceleration, and only observers using the same inertial
frame spatial coordinate with which to measure change of
position $\Delta x$.  We further use the Lorentz-invariant
proper acceleration $a_o$ for describing traveler motion,
so that the relativistic work-energy theorem gives us
$\Delta E \equiv m c^{2} = m a_o \Delta x$, independent of
the time-parameterization in use.  Here m is the rest mass of
our accelerated object, {\em relativistic energy factor}
$\gamma$ is defined dynamically as the accelerated object's
 total energy in units of the rest energy $m c^{2}$, and $c$
is the speed of light.

Consider first the task of assigning to each point on the world
line of an accelerated traveler 3 time parameters, namely $b$,
$t$ and $\tau$, referring respectively to inertial, Newtonian, and
traveler time-maps (or kinematics).  We denote the velocities
corresponding to these time maps, respectively, as $w \equiv dx / db$,
$w \equiv dx / dt$, and $u \equiv dx / d\tau$.  The time maps can
be most simply specified, using the work-energy theorem above, by
determining the dependence of $\gamma$ on the velocities $w$, $v$ and $u$.

{}From the Lorentz expression for $\gamma$ in terms of relativistic
inertial velocity $w$, the $u = \gamma w$ relationship between inertial
and 4-vector velocities, and the familiar Newtonian expression for
kinetic energy $K = \frac12 m v^2$, one can show that
$\gamma \equiv {E / {m c^2}} = \frac{m c^2 + K}{m c^2}
= 1 / \sqrt{1 - {w^2 / c^2}} = 1 + \frac12 {v^2 / c^2}
= \sqrt{1 + {u^2 / c^2}} = {db / {d \tau}}$.  Substitution into the
work-energy equation, $\Delta x = {\Delta \gamma} {c^2 / a_o}$,
yields a {\em velocity-distance equation}, like the familiar
Newtonian $\Delta x = {\Delta (\frac12 v^2)} / a_o$,
for {\em each} of the 3 kinematics.

Taking the time derivative of this displacement equation {\em in
each of the three kinematics} yields a relationship between the
velocities and time and acceleration.  After defining for the
traveler kinematic the {\em hyperbolic velocity angle}
$\eta \equiv asinh (u / c)$ which is additive under Lorentz transform
\cite{Taylor}, and recalling for the inertial kinematic that 4-velocity
$u$ can be expressed in terms of inertial velocity $w$, these
{\em velocity-time equations} can be put into the form
$a_o = {\Delta u} / {\Delta b} = {\Delta v} / {\Delta t} =
c {{\Delta \eta} / {\Delta \tau}}$.  For a fuller comparison, the
equations used for treating constant acceleration, force, and
momentum conservation in introductory physics are summarized in
this way in Table I, while the exact equations for application of
all three kinematics to constant acceleration problems are listed
in Table II.  Note that, except for a correction to the momentum
equation needed at high velocity, namely that momentum $p$ becomes
$m v \sqrt{1 + \frac14 v^2/c^2}$, the introductory physics equations
for one-dimensional motion remain exact at arbitrarily high speed,
even if other velocities and times have to be calculated to provide
information on the kinematic variables (e.g. physical clocks) of interest.

As in introductory physics, armed with a displacement and a velocity equation
for our kinematic and any three of the 5 variables (initial/final velocity,
time elapsed, distance traveled, and constant acceleration), we can solve
for the other two.  Equation {\em pairs} for dealing with this task
for {\em each} of the 10 types of initial condition definable
within {\em each} of the 3 kinematics are listed in Table III.
If the 3 independent variables for a particular problem include
velocities from different kinematics, they can be easily
converted to the kinematic of choice using the expressions above for
$\gamma$ and the fact that $u = v \sqrt{\frac{\gamma + 1}{2}} = w \gamma$.

If time-elapsed values from more than one kinematic are
included in the initial inputs, then the equality above {\em between}
velocity-time expressions can be used, at least numerically, to find
a solution within the kinematic of choice.  Of course, once the
solution for a particular kinematic is in hand, these same equations
allow the 4 velocities and 2 times associated with the other kinematics
to be calculated as well.  Thus for three input variables one now gets
8 quantities of potential interest, instead of only two!

\section{Discussion}

All of the foregoing requires no knowledge of Lorentz transforms, since
the inertial frame for measuring distance traveled was fixed.  Of course,
once such transformations are understood, it is easy to show other things.
For example, since the spatial component of the velocity 4-vector is
$u = c \sinh\eta$, while its time component is $c \gamma = c \cosh\eta$, it
is easy to show that $a_o$ is the magnitude of the 4-vector acceleration
(i.e. the proper time $\tau$ derivative of the 4-vector velocity), and
hence is Lorentz invariant as well as independent of kinematic.

We've also found it to be useful (and enjoyable) in practice to associate
different units with the velocities and times in different kinematics.  In
relativistic problems we are often dealing with distances on the scale of
lightyears, and time on the scale of years, so we often speak of inertial
time $\Delta b$ in "iyears", Newtonian time in year or whatever MKS units are
handy, and traveler time $\Delta \tau$ in "tyears".  Likewise, we usually
speak of inertial velocities $w$ in fractional units of the speed of light
(e.g. $0.5 c$), while leaving Newtonian velocities in whatever units fit the
problem.

Traveler velocities $u$ are most interesting of all, since they exceed the
speed of light even more enthusiastically than Newtonian velocities do
(cf. Fig. 1).  The favorite strategy here, among students I've discussed this
with, has been to define one lightyear per year of traveler velocity as 1
roddenberry or [rb], in tribute to Gene Roddenberry, the producer of the
original StarTrek television series whose offspring continue to ignore the
light speed limit as well as the way that space time is put together at
high speed.  The approach described here might help a bit, in that "warp
speeds" (which still require that one somehow drops out of our space-time
if one is to effect such rapid travel withou inertial effects to passengers
{\em and} "temporal effects" to their colleages at home \cite{Lagoute})
at least can now be mapped to traveler-frame speeds, rather than
to inertial-frame speeds greater than the speed of light (which given the
fabric of spacetime make no sense at all).

We illustrate with an example.  Describe time-elapsed values and final
velocities for a space dragster accelerating from rest at
$a_o = 1 [g] \approx 1.03 [ly/yr^{2}]$
over a distance of 4 lightyears [ly].  Using only the Newtonian equations for
constant acceleration, one can infer (Table III, Row 6, Column 2) that our
dragster will reach a {\em final Newtonian velocity} of $v_f = 2.86 [ly/yr]$
in the {\em Newtonian time} of $\Delta t = 2.79 [yr]$.  The solution is thus
put in hand, and only simple conversion to other kinematics (cf. Table II)
remains!  One obtains {\em inertial time} $\Delta b = 4.87 [iyr]$ using
$u = v \sqrt{1 + \frac14 v^2/c^2}$ and $\Delta u = a_o \Delta b$.  This also
gives {\em final traveler speed} $u_f = 5.03 [rb]$.  One obtains
{\em traveler time} $\Delta \tau = 2.25 [tyr]$ given that
$\eta = {\rm asinh}[u/c]$ and $c {\Delta \eta} = a_o {\Delta \tau}$.
Finally, given $w = u / \gamma$ where $\gamma = 1 + \frac12 v^2 / c^2$,
it follows that the {\em final inertial velocity} $w_f = 0.98 [c]$.

This is only one example.  The rather simple equations of Newtonian constant
acceleration in one dimension allow one to solve a wide array of types of
problems, to which we can now add many relativistic problems as well.
Interactive worksheets and example problems for all of the 30 cases listed in
Table 3 are being assembled on the web at {\bf http://newton.umsl.edu/~run}.

In conclusion, this three-kinematic approach allows one to look at and solve
some relativistic constant-acceleration problems in a new way.  In particular,
it puts both the Newtonian equations for constant acceleration and the
Traveler/Proper Time equations into a new light as fully alternative
relativistic kinematics, even if the former remains self-consistent only
in one spatial dimension.  Moreover, it seems to provide more new answers per
problem (i.e. 6), than there are new concepts required for a post-introductory
physics student to apply and understand them.

\section{Acknowledgements}

Thanks to my colleagues in the Department of Physics and Astronomy
at the University of Missouri in St. Louis, and to the students in
my modern physics classes for both their enthusiasm and their patience.
This paper was prepared with support of the U.S.
Department of Energy Grant No. DE-FG02-92CH10499.

\eject

\end{document}